\documentstyle[PASJadd,epsfig]{PASJ95}

\begin{document}
\setcounter{page}{000}

\title{CO (J=1-0) Observation of the cD Galaxy of AWM7: Constraints on
the Evaporation of Molecular Gas\thanks{OU-TAP 107}}

\author{Yutaka {\sc Fujita}\thanks{JSPS Research Fellow}
\\
{\it Department of Earth and Space Science,
  Graduate School of Science, Osaka University,} 
\\
{\it Machikaneyama-cho,
  Toyonaka, Osaka, 560-0043}
\\[6pt]
Tomoka {\sc Tosaki}, and Akika {\sc Nakamichi}
\\
{\it Gunma Astronomical Observatory, 6860-86, Nakayama, Takayama,
Agatsuma, Gunma, 377-0702} 
\\[6pt] 
and \\ Nario {\sc Kuno}
\\
{\it Nobeyama Radio Observatory, Minamimaki, Minamisaku, Nagano,
384-1305}
}

\abst{We have searched for molecular gas in the cD galaxy of a poor
cluster of galaxies AWM7 using Nobeyama 45 m telescope. We do not detect
CO emission in the galaxy. Our limit of molecular gas in the inner $7.5$
kpc is $M_{\rm H_2}< 4\times 10^8 \MO$. We estimate the total mass of
molecular gas left in the cD galaxy when the gas deposited by a cooling
flow once becomes molecular gas and the molecular gas is continuously
evaporated by the ambient hot gas. The observational limit of molecular
gas requires $f\gtsim 10^{-3}$, where $f$ is the ratio of the heat
conduction rate to that of Spitzer. However, this contradicts recent
X-ray observations showing $f<10^{-5}$. Thus, the non-detection of CO
cannot be explained by the evaporation, and most of the cooled gas
predicted by a cooling flow model may not change into molecular gas in
the cD galaxy. Moreover, we estimate the evaporation time of molecular
clouds brought to a cD galaxy through the capture of gas-rich galaxies
and find that these clouds should not be evaporated if $f\ltsim
10^{-3}-10^{-4}$. Therefore, the non-detection of CO in a cD galaxy
could constrain the total mass of the molecular clouds brought into it.}

\kword{Galaxies: clusters of --- Galaxies: evolution --- Galaxies:
intergalactic medium --- Galaxies: individual (AWM7)}

\maketitle
\thispagestyle{headings}

\section{Introduction}
\label{sec:intro}

The centers of galaxy clusters are usually dominated by very massive
($\sim 10^{13}\;\MO$) galaxies. These galaxies are often called D or cD
galaxies. The observations of cold gas ($\ltsim 10^5$ K) give us the
clues of the formation and evolution of the cD galaxies.

Cold gas in cD galaxies has been investigated from the viewpoint of
cooling flows. The cooling time of intracluster medium (ICM) exceeds the
Hubble time ($\sim 10^{10}$ yr) in the most region of clusters (Sarazin
1986). However, around cD galaxies, the density of ICM increases and the
cooling time decreases to $t_{\rm cool}\ltsim 10^9$ yr. In the absence
of heating, the gas is inferred to be cooling at a rate of $\dot{M}_{\rm
CF}\sim 100\MO\rm\; yr^{-1}$ (Fabian 1994). We will refer to
$\dot{M}_{\rm CF}$ as a mass deposition rate from now on. Thus, the
total mass accumulated around the cD galaxies would result in $\sim
10^{12}\MO$ if the cooling occurred steadily at the rate over the Hubble
time.  Although many observers have tried to detect the cooled gas
mainly in massive cooling flow clusters $\dot{M}_{\rm CF}> 100\;\MO\rm\;
yr^{-1}$, most of them could not detect such a large amount of cold gas.
Using recombination line luminosities, Heckman et al. (1989) estimate
that the total mass of $10^4$ K ionized hydrogen is less than $10^8
\MO$. Observations of the atomic hydrogen 21 cm line limit the total
mass of optically thin H\,{\footnotesize I} to less than $\sim 10^{9}
\MO$ (Burns et al. 1981; Valentijn, Giovanelli 1982; McNamara et
al. 1990). CO observations limit the mass in clouds similar to Galactic
molecular clouds to less than $10^{9-10} \MO$ (McNamara, Jaffe 1994;
O'Dea et al. 1994; Braine, Dupraz 1994). Among the cD galaxies observed
so far ($\sim 30$), only one exception is NGC 1275, the cD galaxy in the
Perseus cluster. The molecular gas of $\sim 10^{9-10}\:\MO$ has been
detected (e.g. Lazareff et al. 1989; Mirabel et al. 1989; Inoue et
al. 1996), although it is smaller than the prediction of the cooling
flow model. These observations may imply that most of the cooled gas
becomes something other than molecular gas such as low mass stars, or
the {\it actual} mass deposition rate, $\dot{M}_{\rm CF}$, may be
reduced by some heating sources.

Before we move to investigate these possibilities, we should consider
another scenario, that is, cooling flows actually exist but the
molecular gas deposited by the cooling flows is continuously evaporated
by the ambient hot ICM. The evaporation time of a molecular cloud is
given by
\begin{equation}
\label{eq:evap_1}
t_{\rm evap}\propto n_{\rm c} R_{\rm c}^2 T_{\rm ICM}^{-5/2}
\:,
\end{equation}
where $n_{\rm c}$ and $R_{\rm c}$ are the density and radius of a
molecular cloud, respectively, and $T_{\rm ICM}$ is the temperature of
hot ICM (Cowie, McKee 1977). In relation (\ref{eq:evap_1}), the
saturation of heat flux is ignored although the following result does
not change significantly even in the saturated case (see \S4). White et
al. (1997) investigate the data of Einstein Observatory and find the
relations $T_{\rm ICM}\propto \dot{M}_{\rm CF}^{0.30}$ and $r_{\rm
cool}\propto \dot{M}_{\rm CF}^{0.25}$, where $r_{\rm cool}$ is the
cooling radius. Thus, we obtain the relation:
\begin{equation}
t_{\rm evap}\propto n_{\rm c} R_{\rm c}^2 
\dot{M}_{\rm CF}^{-0.75}
\end{equation}
On the other hand, if the molecular gas is accumulated by cooling flows
and the age of cooling flows is much larger than $t_{\rm evap}$, the
mass of molecular gas per unit volume is given by
\begin{equation} 
m_{\rm mol}\propto (\dot{M}_{\rm CF}/r_{\rm cool}^3) t_{\rm evap}
           \propto n_{\rm c} R_{\rm c}^2 
           \dot{M}_{\rm CF}^{-0.5}
\end{equation}
Therefore, if $n_{\rm c}$ and $R_{\rm c}$ do not depend on
$\dot{M}_{\rm CF}$ too much, clusters with small $\dot{M}_{\rm CF}$
should have large $m_{\rm mol}$. In these clusters,
we could find molecular gas. 

Since CO has been searched mainly in massive cooling flow clusters
($\dot{M}_{\rm CF}> 100\;\MO\rm\; yr^{-1}$), the observation of clusters
with small $\dot{M}_{\rm CF}$ is important because of the above reason.
Moreover, we note here another importance of searching CO in small
$\dot{M}_{\rm CF}$ clusters. Even if the {\it actual} mass deposition
rate $\dot{M}_{\rm CF}$ is reduced by some heating sources, molecular
gas may be brought to cD galaxies. For example, the capture of gas-rich
galaxies is another possible supply route of molecular gas into cD
galaxies.  In this case, the detection of the molecular gas would be
easier for clusters with small $\dot{M}_{\rm CF}$. This is because the
X-ray emissions of these clusters are weaker than those of clusters with
large $\dot{M}_{\rm CF}$, which means that the heating should be less
effective in these clusters and that the cold gas would be less affected
by the heating.

We search CO in the cD galaxy NGC1129 at the center of a poor cluster
AWM7 with relatively small mass deposition rate ($\dot{M}_{\rm
CF}=41\MO\rm\; yr^{-1}$; Peres et al. 1998). Note that AWM7 is one of
the most closely studied clusters in X-ray. Ezawa et al. (1997) and Xu
et al. (1997) find the metal abundance excess of the ICM at the center
of the cluster. This suggests that the ICM is not well mixed and the
cluster has not experienced violent cluster mergers at least
recently. Thus, the molecular gas in NGC 1129, if exist, would keep
intact.  Moreover, the abundance excess, especially in the cD galaxy (Xu
et al. 1997), implies that there has been star formation activity at the
cluster center. The excess iron mass in the central region ($r<27$ kpc)
is $8\times 10^8\;\MO$. Assuming the 1 $\MO$ iron is ejected into ICM
per 100 $\MO$ of stars formed, the observation shows that $\sim
10^{11}\;\MO$ of stars have been formed in the region. On the other
hand, the present star formation rate of NGC 1129 is $0.04\;\MO\rm\;
yr^{-1}$ within 1.57 kpc form the center (McNamara, O'Connell 1989). If
the distribution of stars in the galaxy is $\propto r^{-2}$, the star
formation rate for $r<27$ kpc is $0.7\;\MO\;\rm yr^{-1}$. Thus, the
present star formation rate in the region is smaller than that the
average through the Hubble time ($\sim 10\;\MO\rm\; yr^{-1}$), and the
star formation in the past must be larger than that at present. If the
`starburst' occurred recently, molecular gas used for it would be left
until present.  In this paper, we assume $H_0=50\;\rm km\;
s^{-1}\;Mpc^{-1}$ throughout.

\section{Observations}
\label{sec:obs}

The $^{12}$CO ($J$ = 1-0) line was observed toward the center of NGC
1129 ($\alpha = 02^{\rm h}51^{\rm m}13^{\rm
s}\hspace{-5pt}.\hspace{2pt}3$, $\delta_{1950} = +41^{\circ}22^{\rm
m}32^{\rm s}$) with the 45-m telescope at Nobeyama Radio Observatory in
1999 March and May.  The half-power beam width (HPBW) was $15''$, which
corresponds to 7.5 kpc at the distance of NGC 1129 ($z = 0.017325$).
The aperture and main beam efficiencies were $\eta_{\rm A}$ = 0.40 and
$\eta_{\rm MB}$ = 0.48, respectively.

We used two SIS receivers that can observe two orthogonal linear
polarizations simultaneously. Martin-Puplett type SSB filters were used
for image sideband rejection. The system noise temperature (SSB)
including the atmospheric effect and the antenna ohmic loss was 400-600
K. As receiver backends 2048-channel wide-band acousto-optical
spectrometers (AOS) were used. The frequency resolution and channel
spacing are 250 kHz and 125 kHz, respectively. Total bandwidth is 250
MHz.  Calibration of the line intensity was made by the chopper-wheel
method, yielding the antenna temperature ($T^{*}_{\rm A}$) corrected for
both atmospheric and antenna ohmic losses. We used the main beam
brightness temperature ($T_{\rm mb} \equiv T^{*}_{\rm A}$/$\eta_{\rm
MB}$) in this paper. The telescope pointing was checked and corrected
every hour by observing the 43GHz SiO maser emission in a late type star
S-Per or W-And. The absolute pointing accuracy was better than $5''$
(peak value) throughout the observations.

\section{Results}
\label{sec:res}

\begin{figure}[t]
\centering \epsfig{figure=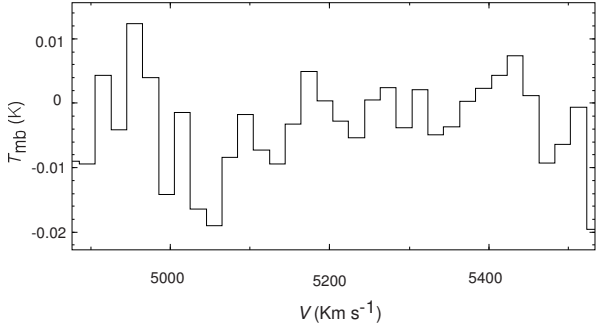, width=8cm} \caption{ Spectrum of NGC
1129 in AWM7 ($z=0.017239$). The velocity resolution is $20\rm\; km\;
s^{-1}$}
\end{figure}
%
%
In figure~1, we present the CO(1-0) spectrum which has been binned by
$20\;\rm km\; s^{-1}$ and has had baseline removed. The spectrum shows
no significant CO(1-0) features either in emission or in absorption.
The $3\sigma$ upper limit to the flux integral is given by
\begin{equation}
I_{\rm CO}= \frac{3\sigma_{\rm ch}\Delta V}
{\sqrt{\Delta V/\Delta V_{\rm ch}}} 
\rm\; K\; km\; s^{-1},
\end{equation}
where $\sigma_c$ is the channel-to-channel rms noise, $\Delta V_{\rm
ch}$ is the smoothed velocity channel spacing, and $\Delta V$ is the
width of line. From figure~1, we obtain $\sigma_{\rm ch}= 0.006$ K.

We assume that the column density of molecular hydrogen is
\begin{equation}
N_{\rm H_2} = 2.8\times 10^{20} I_{\rm CO}\;\rm cm^{-2}
\end{equation}
(O'Dea et al. 1994). The total mass of molecular hydrogen is given by
\begin{equation}
\label{eq:mol}
M_{\rm mol} = \frac{\pi r^2}{4\ln 2}N_{\rm H_2}m_{\rm H_2}\;,
\end{equation}
where $r$ is the beam size at the distance of the source and $m_{\rm
H_2}$ is the mass of the hydrogen molecule (O'Dea et al. 1994). We
assume that $\Delta V = 300\;\rm km\; s^{-1}$ for a rectangular line
feature. This is the same as McNamara and Jaffe (1994) and nearly
corresponds to the internal velocity dispersion of NGC 1129 (McElroy
1995). From equation (\ref{eq:mol}), we obtain $M_{\rm mol}<4\times
10^{8}\;\MO$ within 7.5 kpc from the center. This is one of the most
sensitive limits for cD galaxies.

\section{Discussion}


Although $\dot{M}_{\rm CF}$ of AWM7 is relatively small, the
non-detection of molecular gas conflicts with a cooling flow model if
the cooled gas becomes molecular gas and if we ignore the effect of the
evaporation. Peres et al. (1998) estimate that the mass deposition rate
and cooling radius of AWM7 are $\dot{M}_{\rm CF}=41\MO\;\rm yr^{-1}$ and
$r_{\rm cool}=103$ kpc, respectively. The analysis based on a cooling
flow model shows that the mass deposited within $r$ is
$\dot{M}(<r)=\dot{M}_{\rm CF}(r/r_{\rm cool})$ (Fabian 1994). Thus, the
mass deposition rate within the beam of Nobeyama 45m is $\dot{m}=3.8
\MO\rm\; yr^{-1}$, considering the projection effect. Thus, molecular
gas of $\sim 10^{10}\MO$ would be detected if the cooling flow occurred
steadily at the rate over the Hubble time.

As mentioned in \S1, when the molecular gas deposited by a cooling flow
is continuously evaporated by the ambient hot ICM, the detection of
molecular gas would be relatively easy in clusters with small
$\dot{M}_{\rm CF}$. Although we cannot detect CO, it constrains the
evaporation rate of molecular gas and the heat conduction rate of the
ICM. Using the results, we could investigate whether the non-detection
of CO is consistent with the evaporation model.

The accumulation time of molecular gas is
\begin{equation}
\label{eq:tacc}
t_{\rm acc}=M_{\rm
mol}/\dot{m}\ltsim 1\times 10^8\rm\; yr \;.
\end{equation}
Although the properties of the molecular clouds deposited by cooling
flows are not well-known, we could calculate the evaporation time of the
clouds as follows. In disk galaxies, molecular gas is considered to be
produced through disk instabilities (e.g. Larson 1987). In elliptical
galaxies like cDs, we expect that the mechanism is ineffective. Instead,
we expect that the molecular gas is produced through the thermal
instability of ICM. One possible seed of the instability is supernova
remnants (Fujita et al. 1996, 1997). Thus, we assume that a supernova
remnant is the seed of a molecular cloud and that only thermal
evaporation affects the cloud after the formation, although these may
oversimplify the evolution of molecular clouds (refer to Loewenstein and
Fabian [1990] for more realistic discussion about the issue). Note that
the results in the following can be applied to other formation
mechanisms of cloud if the resultant mass is nearly the same. If we can
ignore the fragmentation and coalescence of molecular clouds, the mass
of a molecular cloud is equal to that of a supernova remnant. Since the
radius of a supernova remnant is given by
\begin{equation}
\label{eq:snr}
R_{\rm s}\sim 
50\;{\rm pc}\left(\frac{P_{\rm ICM}}{4\times 
10^5\:\rm cm^{-3}\;K}\right)^{-1/3}
\left(\frac{E_{\rm SN}}{10^{51}\rm\;ergs}\right)^{1/3}
\end{equation}
(Fujita et al. 1997), the mass of a molecular cloud is
\begin{eqnarray}
M_{\rm c}&=&\frac{4}{3}\pi R_{\rm s}^3 m_{\rm H} n_{\rm ICM}\\
         \nonumber
         &=&130\;\MO
            \left(\frac{n_{\rm ICM}}{10^{-2}\;\rm cm^{-2}}\right)\\
        & &\times
            \left(\frac{P_{\rm ICM}}{4\times 
         10^5\:\rm cm^{-3}\;K}\right)^{-1}
	    \left(\frac{E_{\rm SN}}{10^{51}\rm\;ergs}\right)
\label{eq:mass}\;,
\end{eqnarray}
where $P_{\rm ICM}$ and $n_{\rm ICM}$ are the pressure and the density of
ICM, respectively, $E_{\rm SN}$ is the energy released by a supernova,
and $m_{\rm H}$ is the hydrogen mass.

If molecular clouds are in pressure equilibrium with the ambient ICM,
the density of the molecular gas is 
\begin{equation}
\label{eq:nc}
n_{\rm c} = 4\times 10^4\;{\rm cm^{-3}}
\left(\frac{P_{\rm ICM}}{4\times 10^5\:\rm cm^{-3}\;K}\right)
\left(\frac{T_{\rm c}}{10\rm\; K}\right)^{-1}
      \;,
\end{equation}
where $T_{\rm c}$ is the temperature of the molecular gas. Since $n_{\rm
ICM}m_{\rm H}R_{\rm s}^3=n_{\rm c}m_{\rm H_2}R_{\rm c}^3$, the radius of
a molecular cloud is
\begin{eqnarray}
R_{\rm c}&\sim& 0.25\;{\rm pc}
      \left(\frac{R_{\rm s}}{50\;\rm pc}\right)
      \left(\frac{T_{\rm c}}{10\;\rm K}\right)^{1/3}
\nonumber\\
 & & \times
      \left(\frac{T_{\rm ICM}}{4\times 10^7\;\rm K}\right)^{-1/3}
      \label{eq:Rc} \;
\end{eqnarray}

We assume that heat is supplied to molecular clouds from the isothermal
X-ray gas component prevailing even in the central region of clusters
(e.g. Ikebe et al. 1999). Moreover, we assume that the isothermality of
the component is retained by adiabatic heating or magnetic loops
connected to the overall thermal reservoir of the cluster (e.g. Norman,
Meiksin 1996) or the large heat conduction rate of the gas constituting
the component. The evaporation time of a molecular cloud embedded in ICM
is given by
\begin{eqnarray}
\label{eq:evap}
t_{\rm evap} &=& 1\times 10^{5}\;{\rm yr}\;
\frac{1}{f}
\left(\frac{n_{\rm c}}{4\times 10^4\rm\; cm^{-3}}\right)\nonumber\\
 & & \times
\left(\frac{R_{\rm c}}{0.25\rm\; pc}\right)^2 
\left(\frac{T_{\rm ICM}}{4\times 10^7\rm\;K}\right)^{-5/2} \\
&=& 1\times 10^{5}\;{\rm yr}\; 
\frac{1}{f}
\left(\frac{n_{\rm c}}{4\times 10^4\rm\; cm^{-3}}\right)^{1/3}\nonumber\\
 & & \times
\left(\frac{M_{\rm c}}{130\; \MO}\right)^{2/3} 
\left(\frac{T_{\rm ICM}}{4\times 10^7\rm\;K}\right)^{-5/2} 
\label{eq:evap_mc}
\end{eqnarray}
(Cowie, McKee 1977). Equation (\ref{eq:evap_mc}) shows that if $M_{\rm
c}$ is larger than that given in equation (\ref{eq:mass}), $t_{\rm
evap}$ should be larger. The parameter $f$ is the ratio of the heat
conduction rate to that of Spitzer (1962) and $0\leq f \leq 1$ (Cowie
and McKee [1977] assume $f=1$). When $f<1$, the heat conduction rate and
the mean free path of an electron are considered to be regulated by
plasma instabilities around the cloud (Pistinner et al. 1996; Hattori,
Umetsu 1999). If the mean free path of an electron is comparable or even
greater than the radius of a cloud, the thermal evaporation is
saturated. Defining the saturation parameter,
\begin{eqnarray}
\label{eq:sat}
\sigma_0&=&2700\;f\left(\frac{T_{\rm ICM}}{4\times 10^7\;\rm K}\right)^2
	   \nonumber \\
           & & \times
           \left(\frac{n_{\rm ICM}}{10^{-2}\rm\; cm^{-3}}\right)^{-1}
	   \left(\frac{R_{\rm c}}{0.25\rm\; pc}\right)^{-1}
	   \;,
\end{eqnarray}
the saturation occurs when $\sigma_0\gtsim 1$ (Cowie and McKee 1977).
In the saturated case, the evaporation time is given by
\begin{eqnarray}
\label{eq:evap_sat}
t_{\rm evap} &=& 8\times 10^{6}\;{\rm yr}\;
\left(\frac{n_{\rm c}}{4\times 10^4\rm\; cm^{-3}}\right)\nonumber\\
 & & \times
\left(\frac{n_{\rm ICM}}{10^{-2}\rm\; cm^{-3}}\right)^{-1}
\left(\frac{R_{\rm c}}{0.25\rm\; pc}\right) 
\nonumber\\
 & & \times
\left(\frac{T_{\rm ICM}}{4\times 10^7\rm\;K}\right)^{-1/2}
\left(\frac{\sigma_0}{2700}\right)^{-3/8} \\
&=& 8\times 10^{6}\;{\rm yr}\;
\left(\frac{n_{\rm c}}{4\times 10^4\rm\; cm^{-3}}\right)^{2/3}\nonumber\\
 & & \times
\left(\frac{n_{\rm ICM}}{10^{-2}\rm\; cm^{-3}}\right)^{-1}
\left(\frac{M_{\rm c}}{130\; \MO}\right)^{1/3} 
\nonumber\\
 & & \times
\left(\frac{T_{\rm ICM}}{4\times 10^7\rm\;K}\right)^{-1/2}
\left(\frac{\sigma_0}{2700}\right)^{-3/8}\label{eq:evap_mcs}
\end{eqnarray}
(Cowie, McKee 1977).

We will adopt $T_{\rm c}=10\rm\; K$ and $E_{\rm SN}=10^{51}\;\rm ergs$;
the temperature of a molecular cloud is the typical one in the Galaxy
(Scoville, Sanders 1987). If we adopt the observed values $n_{\rm
ICM}=1.82\times 10^{-2}\rm\; cm^{-3}$ and $T_{\rm ICM}=3.63$ keV (Mohr
et al. 1999; Ezawa et al. 1997), equation (\ref{eq:snr}) and
(\ref{eq:Rc}) yield $R_{\rm c}=0.20$ pc and equation (\ref{eq:sat})
shows that the saturation occurs when $f\gtsim 5\times 10^{-4}$. If the
age of a cooling flow ($t_0\sim 10^{10}$ yr; Fabian 1994) is much larger
than the evaporation time of a molecular cloud, the evaporation of
molecular gas should be balanced with the accumulation. In this case,
the evaporation time should be equal to the accumulation time. From
equations (\ref{eq:tacc}), (\ref{eq:evap}), (\ref{eq:sat}), and
(\ref{eq:evap_sat}), this requires $f\gtsim 10^{-3}$. However, recent
observations of ASCA show that ICM is inhomogeneous in temperature at
least in some clusters (e.g. Ikebe et al. 1999). In order to explain
this inhomogeneity by the cooling flow model, $f$ must be less than
$10^{-5}$ at least around cooler X-ray gas (Pistinner et al. 1996;
Hattori, Umetsu 1999) if the cooler gas component is not isolated by
something like a magnetic field.  Therefore, as long as $f<10^{-5}$, the
evaporation cannot account for the non-detection of CO and most of the
cooled gas may become something other than molecular gas such as dust
(Fabian et al. 1994; Voit, Donahue 1995; Edge et al. 1999) or low mass
stars (Sarazin, O'Connell 1983), or there may be something wrong in the
cooling flow model, that is, the actual mass deposition rate is much
less than the one estimated by X-ray observations.

So far, we have not considered the molecular gas brought by gas-rich
galaxies merged into cD galaxies including NGC1129. Finally, we examine
the evaporation of this kind of molecular gas. The mass of a molecular
cloud brought through the capture would be larger in comparison with the
case of cooling flows ($\sim 100\;\MO$; equation [\ref{eq:mass}]). If
molecular clouds in captured galaxies are similar to those in our
Galaxy, the masses are typically $M_{\rm c}\gtsim 10^5\;\MO$ (Binney,
Tremaine 1987). Thus, if we adopt equation (\ref{eq:nc}) and the
normalizations therein, $R_{\rm c}\gtsim 2.5$ pc. From equations
(\ref{eq:sat}) and (\ref{eq:evap_mc}), we expect $t_{\rm evap}\sim
10^{12}$ yr for $f=10^{-5}$.  Moreover, even if $f=10^{-3}$, we obtain
$t_{\rm evap}\sim 10^{10}$ yr. Thus, the evaporation can be
ignored. This means that if molecular gas is brought into a cD galaxy
through galaxy captures and $f\ltsim 10^{-3}$, the gas should be left
there. (It is to be noted that when $T_{\rm ICM}\sim 10$ keV, equation
[\ref{eq:evap_mc}] shows that the condition $t_{\rm evap}\gtsim 10^{10}$
requires $f\ltsim 10^{-4}$.) Hence, the non-detection of molecular gas
strongly constrains the amount of molecular gas brought into the cD
galaxy through galaxy captures. Using a theoretical model based on a
hierarchical clustering scenario, Fujita et al. (1999) predict the
amount of the molecular gas and compare it with the observations.

\section{Conclusions}

We have searched for CO emission from the cD galaxy NGC 1129 in AWM7. We
have obtained the upper limit of molecular hydrogen mass ($4\times
10^8\;\MO$). This is one of the most sensitive limits for cD
galaxies. We predict the total mass of molecular gas left in the cD
galaxy on the assumption that while the gas deposited by a cooling flow
once becomes molecular gas, the molecular gas is continuously evaporated
by the ambient hot gas. We find that the upper limit of molecular
hydrogen mass shows $f\gtsim 10^{-3}$, where $f$ is the ratio of the
heat conduction rate to that of Spitzer (1962). However, this is
inconsistent with recent X-ray observations showing $f<10^{-5}$. Thus,
most of the cooled gas predicted by a cooling flow model does not seem
to become molecular gas in the cD galaxy. Therefore, if $f<<1$ as is
suggested by the X-ray observations, the ultimate fate of most of the
cooled gas may be something other than molecular gas such as dust or low
mass stars. Alternatively, the actual mass deposition rate may be much
less than the one predicted by a cooling flow model.

We find that molecular clouds brought to a cD galaxy by the gas-rich
galaxies captured by the cD should not be evaporated when $f\ltsim
10^{-3}-10^{-4}$. This implies that if we obtain the upper limit of the
mass of molecular gas in a cD galaxy, we could constrain the supply of
molecular gas brought into through the galaxy captures.

\par
\vspace{1pc}\par
We thank an anonymous referee for invaluable advice and suggestions.
This work was supported in part by the JSPS Research Fellowship for
Young Scientists. 

\section*{References}
\small




\re Binney J., Tremaine S.\ 1987, Galactic Dinamics (Princeton; New
Jersey)

\re Braine J., Dupraz C.\ 1994, A\&A 283, 407


\re Burns J.O., White R.A., Haynes M.P.\ 1981, AJ 86, 1120

\re Cowie L.L., McKee C.F.\ 1977, ApJ 211, 135


\re Edge A.C., Ivison R.J., Smail I., Blain A.W.,Kneib J.-P.\ 1999,
MNRAS 306, 599


\re Ezawa H., Fukazawa Y., Makishima K., Ohashi T., Takahara F., Xu H., 
Yamasaki N.Y.\ 1997, ApJL, 490, 33

\re Fabian A.C.\ 1994, ARA\&A 32, 77

\re Fabian A.C., Johnstone R.M., Daines S.J.\ 1994, MNRAS 271, 737 

\re Fujita Y., Fukumoto J., Okoshi, K.\ 1996, ApJ 470, 762

\re Fujita Y., Fukumoto J., Okoshi, K.\ 1997, ApJ 488, 585

\re Fujita Y., Nagashima M., Gouda N.\ 1999, PASJ submitted


\re Heckman T.M., Baum S.A., van Breugel W.J.M., McCarthy P.J.\ 1989,
ApJ 338, 48

\re Hattori M, Umetsu K.\ 1999, ApJ in press


\re Ikebe Y., Makishima K., Fukazawa Y., Tamura T., Xu H., Ohashi T.,
Matsushita K.\ 1999, ApJ 525, 58

\re Inoue M.Y., Kamono S., Kawabe R., Inoue M., Hasegawa T., Tanaka M.\
1996, AJ 1111, 1852

%
%
%

\re Larson R.B.\ 1987, in Starbursts and Galaxy Formation, ed. Trinh
Xuan Thuan, T. Montmerle, and J. Tran Thanh Van (Editions Frontieres;
France)

\re Lazareff B., Castets A., Kim D.W., Jura M.\ 1989, ApJL 336 13

\re Loewenstein M., Fabian A.C.\ 1990, MNRAS, 242, 120


\re Mirabel I.F., Sanders D.B., Kazes I.\ 1989, ApJL 340, 9

\re McElroy D.B.\ 1995, ApJS 100, 105

\re McNamara B.R., Bregman J.N., O'Connell R.W.\ 1990, ApJ 360, 20

\re McNamara B.R., Jaffe W.\ 1994, A\&A 281, 673

\re McNamara B.R., O'Connell R.W.\ 1989, AJ 98, 2018 

%

\re Norman C., Meiksin A.\ 1996, ApJ 468, 97 

\re Mohr J.J., Mathiesen B., Evrard A.E.\ 1999, ApJ 517, 627

\re O'Dea C.P., Baum S.T., Maloney P.R., Tacconi L.J., Sparks W.B.\
1994, ApJ 422, 467

\re Pistinner S., Levinson A., Eichler D.\ 1996, ApJ 467, 162


\re Peres C.B., Fabian A.C., Edge A.C., Allen S.W., Johnstone R.M.,
White D.A.\ 1998, MNRAS 298, 416


\re Sarazin C.L.\ 1986, Phys. Mod. Rev. 58, 1.

\re Sarazin C.L., O'Connell R.W.\ 1983, ApJ 268, 552

\re Scoville N.Z., Sanders D.B.\ 1987, Interstellar Processes, p21
(Tokyo:Dordrecht)

\re Spitzer L.\ 1962, Physics of Fully Ionized Gases (New York:
Interscience)




\re Valentjin E.A., Giovanelli R.\ 1982, A\&A, 114, 208

\re Voit G.M., Donahue M.\ 1995, ApJ, 452, 164 


\re White D.A., Jones C., Forman W.\ MNRAS 292, 419


\re Xu H., Ezawa H., Fukazawa Y., Kikuchi K., Makishima K., Ohashi T.,
Tamura T.\ 1997, PASJ 49, 9


%

\end{document}